\documentstyle[preprint,aps]{revtex}
\begin{document}

\title{Photoluminescence signature of skyrmions at $\nu = 1$}

\draft

\author{T. Portengen, J. R. Chapman, V. Nikos Nicopoulos, and N. F. Johnson}
\address{Department of Physics, University of Oxford,
Parks Road, Oxford OX1 3PU, United Kingdom}

\date{\today}

\maketitle

\begin{abstract}
The photoluminescence spectrum of quantized Hall states
near filling factor $\nu=1$ is investigated theoretically. 
For $\nu \geq 1$ the spectrum consists of a right-circularly 
polarized (RCP) line and a left-circularly polarized (LCP) line, 
whose mean energy: (1) does not depend on the 
electron $g$ factor for spin-$\frac{1}{2}$ quasielectrons, 
(2) does depend on $g$ for charged spin-texture excitations 
(skyrmions). For $\nu < 1$ the spectrum consists of a LCP line 
shifted down in energy from the LCP line at $\nu \geq 1$. 
The $g$\/-factor dependence of the red shift of the LCP line 
determines the nature of the negatively charged excitations.  
\end{abstract}


   The study of the quantum Hall effect has recently focused 
    on filling factors near $\nu = 1$. At $\nu = 1$ the Fermi 
   level is located between the lowest spin-split Landau levels, 
    yielding a fully spin-polarized ground state of
    the two-dimensional electron system (2DES). The nature of the 
      elementary charged excitations is determined by the 
     interplay of the Zeeman energy $\frac{1}{2}g_{e}\mu_{B}B$ 
     and the exchange energy $e^{2}/\epsilon \ell$, where
    $g_{e}$ is the electron $g$ factor ($g_{e} = -0.44$ in GaAs), 
     $\ell = \sqrt{\hbar c/eB}$ is the magnetic length, $\mu_{B}$ 
       is the Bohr magneton, 
     and $\epsilon$ is the background dielectric constant. For 
    small ($<0.02$) values of the parameter $g=\frac{1}{2}|g_{e}|\mu_{B}B
      /(e^{2}/\epsilon \ell)$ the lowest-energy negatively 
    (positively) charged excitations are not spin-$\frac{1}{2}$ 
        quasielectrons 
     (quasiholes), but objects of much larger spin 
      called skyrmions~\cite{Sondhi} 
       (antiskyrmions) or charged spin-texture excitations~\cite{Fertig}.
     The gradual spin reversal that characterizes skyrmions increases 
      the Zeeman energy, but reduces the exchange energy due to the near
     parallelism of neighbouring spins. The spin-$\frac{1}{2}$ 
    quasielectrons
  can be regarded as spin textures with zero radius~\cite{Fertig}. 
     Experimental 
    evidence for skyrmion excitations at $\nu = 1$ has been obtained in  
  NMR Knight-shift studies~\cite{Barrett}, tilted-field magnetotransport 
    measurements~\cite{Schmeller}, and polarized interband optical 
   transmission spectroscopy~\cite{Aifer}. Recent magneto-photoluminescence 
       experiments have studied the optical recombination of 
       2D electrons with itinerant holes in single 
       heterojunctions~\cite{Turberfield}, 
     with holes confined to the undoped side of a one-side doped 
      quantum well~\cite{Goldberg}, and with holes bound to neutral 
        acceptors in Be $\delta$-doped structures~\cite{Buhmann}.
     A clear indication of skyrmions in the photoluminescence 
       (PL) spectrum at $\nu = 1$ has thus far been lacking. 

   Here we predict a PL signature of skyrmions at $\nu = 1$. 
      The energy of the LCP line exhibits a red shift at $\nu = 1$ 
    as the magnetic field is increased. The magnitude of the red
    shift does not depend on $g$ if the elementary negatively
    charged excitations are quasielectrons, but does
    depend on $g$ if they are skyrmions. 
    Measurements of the variation of the red shift with tilting 
     angle or applied hydrostatic pressure allow the detection 
    of skyrmions using PL spectroscopy.
   
    We consider the interband recombination of electrons 
     in the ground subband of the confining potential  
    with holes in the GaAs valence band, in the presence of 
     a strong $B$ field along the $z$ direction.
   In the vicinity of $\nu = 1$, transitions are 
    observed~\cite{Goldberg} between the energy levels shown 
  in the inset of Fig.~\ref{fig:mean}.
  The main LCP transition   
   is between the $m_{s} = +\frac{1}{2}$ (spin-up)
      electron lowest Landau level (LLL) 
     and the $m_{j} = +\frac{3}{2}$ heavy-hole 
      LLL, and the main RCP transition is between the 
     $m_{s} = -\frac{1}{2}$ (spin-down) electron LLL and the 
     $m_{j} = -\frac{3}{2}$ heavy-hole LLL.\@ 
   Since the PL signature of skyrmions is independent of the 
      relative intensities of the LCP and RCP
       lines, we can for simplicity assume equal populations 
       of the $m_{j} = +\frac{3}{2}$ and $m_{j} = -\frac{3}{2}$ 
     hole states prior to recombination. 

We first examine recombination exactly at $\nu = 1$. The ground 
    state of the $N$\/-electron system before photoexcitation 
     is the filled spin-up LLL.\@ We assume
    that following photoexcitation, the $(N+1)$\/-electron
     system relaxes to its ground state in the presence of the
     valence hole on a time scale that is short compared to
      the radiative recombination time. The Hamiltonian 
  for $(N+1)$ electrons in the LLL interacting with a single 
    valence hole in the LLL is
\begin{eqnarray}
    H  & = & \frac{1}{2} g_{e}\mu_{B} B \sum_{m \sigma} \sigma \,
    e^{\dagger}_{m \sigma} e_{m \sigma} +
      \sum_{m \sigma} \varepsilon_{h \sigma} 
    h^{\dagger}_{m \sigma} h_{m \sigma}     
\nonumber \\ 
       &  &  +       
     \frac{1}{2}  \sum_{\sigma \sigma'} \sum_{m m' m'' m'''}
 V^{ee}_{m m' m'' m'''} e^{\dagger}_{m \sigma}
    e^{\dagger}_{m' \sigma'} e_{m'' \sigma'} e_{m''' \sigma}
\nonumber  \\
       &  & -   
      \sum_{\sigma \sigma'}  \sum_{m m' m'' m'''}
      V^{eh}_{m m' m'' m'''} e^{\dagger}_{m \sigma} 
        h^{\dagger}_{m' \sigma'} h_{m'' \sigma'} e_{m''' \sigma} .
\end{eqnarray}
   Here $e^{\dagger}_{m \sigma}$ creates an electron with 
    $m_{s} = \frac{1}{2} \sigma$ ($\sigma = \pm 1$) in the 
     state $\phi_{m}({\bf r}) =  
  (2^{m+1} \pi\, m!)^{-1/2} r^{m} e^{-i m \phi} e^{-r^{2}/4}$ 
   ($\ell = 1$), and $h^{\dagger}_{m \sigma}$ creates a hole with 
  $m_{j} = \frac{3}{2} \sigma$ in the state $\phi^{*}_{m}({\bf r})$.
    A uniform neutralizing background 
      is added to the Hamiltonian in the usual way.
   The energies of the hole states are 
\begin{equation}
\label{eq:hole}
    \varepsilon_{h\,\sigma} = E_{g} + \frac{eB}{2 \mu c}
     - \frac{3}{2} \sigma g_{h} \mu_{B}B ,
\end{equation}
     where we have taken the Fermi level at $\nu = 1$ as
      the zero of our energy scale.
   Here $E_{g}$ is the gap between the ground electron 
     subband and the GaAs valence band, $\mu = m_{e}m_{h}/
     (m_{e}+m_{h})$ is the reduced mass, 
      $m_{e}$ and $m_{h}$ are the electron and hole 
    in-plane effective masses, and $g_{h}$ is the hole 
      $g$ factor. $V^{ee}_{m m' m'' m'''}$ and 
    $V^{eh}_{m m' m'' m'''}$ are matrix elements of the 
    electron-electron interaction, 
     $V^{ee}({\bf r}) = e^{2}/\epsilon |{\bf r}|$,
   and the electron-hole interaction, $V^{eh}({\bf r}) = 
   e^{2}/\epsilon |{\bf r}+\hat{z}d|$.
    Here $d$ is the distance between the planes to which
   the electrons and the hole are confined, and $\hat{z}$ is
 a unit vector along $z$. The confinement of
  the hole along $z$ occurs naturally in 
    a one-side doped quantum well, and is also expected
   in a single heterojunction~\cite{Birman}. 
       
 The initial state $|i_{\sigma}\rangle$ prior to recombination 
  is the ground state of the $(N+1)$\/-electron system in the 
 presence of a $m_{j} = -\frac{3}{2} \sigma$ valence hole. 
    Here $\sigma = +1$ for RCP, and $\sigma = -1$ for LCP.
   Consider recombination in the presence of disorder. The 
    disorder may be caused, for example, by interface roughness 
   or impurities in the GaAs layer. The disorder is likely to 
     localize the hole in a potential minimum, whose location 
    we choose as the origin of our coordinate system.
   Provided the potential varies slowly on the scale of $\ell$, 
      we may take the hole to be in the $m = 0$ state.
   The ground state of the $(N+1)$\/-electron system in the 
      absence of the hole is known to be a skyrmion~\cite{Sondhi}.
   This leads us to consider initial states of the 
     form~\cite{Fertig}
\begin{equation}
\label{eq:state}
 | i_{\sigma} \rangle = \prod_{m = -1}^{\infty} 
   ( -u_{m} e^{\dagger}_{m+1 \downarrow}
    + v_{m} e^{\dagger}_{m \uparrow} ) 
            h^{\dagger}_{0,-\sigma} |{\rm vac}\rangle ,
\end{equation}
   where $u_{-1} = -1$ and $v_{-1} = 0$. Here $|{\rm vac}\rangle$
     is the vacuum state with a filled valence band and empty
         conduction band. The state $|i_{\sigma}\rangle$ describes 
   a skyrmion bound to a localized hole. The parameters 
       $u_{m}$ and $v_{m}$ for $m \geq 0$ are determined by 
     minimizing $\langle i_{\sigma}|H| i_{\sigma}\rangle$. The  
   recombination in the absence of disorder is  
      discussed below. The PL signature of skyrmions 
   in the absence of disorder is qualitatively similar to that 
      in the presence of disorder. 
   
  The initial states considered in a previous theoretical study 
  of photoluminescence at $\nu = 1$ were excitonic states 
  consisting of a filled spin-up LLL and a spin-down electron 
  bound to a valence hole~\cite{Cooper}. In the context of
   the present model, the excitonic states correspond to a 
   filled spin-up LLL and a spin-down electron bound to 
   a hole in the $m = 0$ state.
   These states are obtained by setting $u_{m} = 0$ and 
   $v_{m} = 1$ for $m \geq 0$ in Eq.~(\ref{eq:state}).
     The excitonic states do not allow for the formation of
  a spin texture of nonzero size in the presence of the 
     hole. Hence no general PL signature of skyrmions was
    obtained in Ref.~\onlinecite{Cooper}. 

   The initial states considered in this work {\em do} 
     allow spin textures to form in the presence of the 
   hole. The electron-hole interaction 
    reduces the radius of the spin texture.
    We have calculated numerically the magnitude of the 
     spin $S_{z}$ of the spin texture as a function of 
    the distance $d$ between the electron and hole planes.
   We find that when $d > \ell$ the ground state corresponds
    to a spin texture with finite radius 
     ($|S_{z}| > \frac{1}{2}$), and when $d < \ell$ the 
    ground state corresponds to a spin texture with zero 
    radius ($|S_{z}| = \frac{1}{2}$). Our results agree
     with those obtained by other 
        workers~\cite{Cooper,Brey}. 
    As an example, consider a GaAs-AlGaAs heterojunction
   with $n_{s} = 10^{11}$~cm$^{-2}$. The magnetic
 length at $\nu = 1$ is $\ell = 126$~${\rm \AA}$. We find 
    $|S_{z}| = 4.58$, $2.94$, $1.92$, $0.5$ for 
   $d = \infty$, $3 \ell$, $2 \ell$, $\ell$. 
     The wide quantum wells in 
    Ref.~\onlinecite{Goldberg} have well widths of 
    $400$~${\rm \AA}$ and $500$~${\rm \AA}$ with 
    $n_{s} = 2.8 \times 10^{11}$~cm$^{-2}$ and 
    $n_{s} = 1.9 \times 10^{11}$~cm$^{-2}$, respectively. 
     Based on our numerical results, we therefore expect a 
        PL signature of skyrmions in such systems.
   
   We treat the optical recombination in the electric-dipole
      approximation. The dipole matrix element between the
    Bloch wavefunctions of the GaAs conduction band and valence 
  band gives rise to the selection rule $M'_{J} = M_{J} - \sigma$, 
    where $M_{J}$ and $M'_{J}$ are the $z$ components of the total 
   angular momentum  before and after the transition. The selection
    rule for the total angular momentum replaces the usual selection
    rules for the orbital and spin angular momenta due to the 
  spin-orbit coupling of the GaAs valence band. The $z$ component 
   of the total spin of the $N$\/-electron state 
     $|f_{\sigma} \rangle$ after recombination 
       must satisfy the spin selection rule 
\begin{equation}
\label{eq:rule}
  M'_{S} = \frac{1}{2} N + S_{z} + \frac{1}{2} \sigma.
\end{equation}
     This selection rule leads to an important 
     difference between the recombination of a 
      quasielectron and the recombination of a skyrmion.
     While recombination of a quasielectron leaves 
      either zero (in the case of RCP) or one (in the case 
     of LCP) spin flip in the final state, recombination of a
    skyrmion leaves a large number of spin flips in the final 
      state. The number of spin flips left is 
         $|S_{z} + \frac{1}{2} \sigma|$. 

    The PL spectrum is 
\begin{equation}
    P_{\sigma}(\omega) = 2\pi \sum_{f} 
    |\langle f_{\sigma} | L_{\sigma} | i_{\sigma} \rangle|^{2} 
         \delta(E_{i} - E_{f} - \omega) , 
\end{equation}
    where $L_{\sigma} = \mu_{\sigma} \sum_{m} 
      e_{m,-\sigma} h_{m,-\sigma}$ is the 
   luminescence operator, and $E_{i}$ and $E_{f}$ are 
     the energies of the initial and final states. 
    Here $\mu_{\sigma}$ is the product of the interband 
    dipole matrix element and the overlap between the
   electron and hole $z$ wavefunctions.
    The PL signature of skyrmions occurs in the luminescence 
     energies. We obtain expressions for the moments $\langle 
    \omega^{n}_{\sigma} \rangle = \int\! d\omega\,
    \omega^{n} P_{\sigma}(\omega)$ of the 
    luminescence lines by summing over a complete set of 
   final states within the LLL.\@ Following the same
    algebraic steps as in the derivation of the 
      sum rules for the whole PL spectrum~\cite{Bethe},
     we find
\begin{equation}
\label{eq:moments}
      \langle \omega^{n}_{\sigma} \rangle =  
    \langle i_{\sigma} |L^{\dagger}_{\sigma}
    [ L_{\sigma},H ]_{n}|i_{\sigma} \rangle ,
\end{equation}
    where
  $[L_{\sigma},H]_{n} = [[L_{\sigma},H]_{n-1},H]$, and 
  $[L_{\sigma},H]_{0} = L_{\sigma}$. The moments 
    do not obey the sum rules for the whole PL spectrum 
     due to the 
    projection of $L_{\sigma}$ and $H$ onto the LLL.\@ 
  The usefulness of moments was shown by
  Apalkov and Rashba in the context of the fractional quantum 
  Hall effect~\cite{Rashba}. It is convenient to redefine
    $\langle \omega^{n}_{\sigma} \rangle \equiv  
     \langle \omega^{n}_{\sigma} \rangle /   
     \langle \omega^{0}_{\sigma} \rangle$.    

    We calculate the energies of the
  RCP and LCP lines by evaluating Eq.~(\ref{eq:moments})
    with $n = 0,1$ in the initial state given by 
   Eq.~(\ref{eq:state}). Three terms contribute
     to $\langle \omega_{\sigma} \rangle$. The first 
    term is $\varepsilon_{h,-\sigma}$.
      The second term is the energy cost 
   to remove an electron with $m = 0$ and 
    $m_{s} = -\frac{1}{2} \sigma$ from the spin texture.
        The third term is the energy of the state 
     $L_{\sigma}|i_{\sigma}\rangle$ in the potential
      $-V^{eh}({\bf r})$ switched on by the removal 
     of the hole. Figure~\ref{fig:mean} shows 
  the mean energy of the RCP and LCP lines,
\begin{equation}
    \langle \omega \rangle = \frac{ \langle \omega_{+} \rangle 
  + \langle \omega_{-} \rangle}{2} ,
\end{equation}
   as a function of $g$, for various values of $d$. 
  The $g$ dependence of $\langle \omega \rangle$ 
   is a PL signature of skyrmions.  
  By setting $u_{m} = 0$ and $v_{m} = 1$ for $m \geq 0$ 
    it can be shown that $\langle \omega \rangle$  
 {\em does not} depend on $g$ when the initial state consists 
   of a quasielectron and a hole.  
   $\langle \omega \rangle$ {\em does} depend on $g$ 
   when the initial state consists of a skyrmion and a 
    hole. The value of $g$ below which skyrmions exist in 
   the initial state decreases as the hole approaches
    the electron plane.
              
    We now consider the PL spectrum 
     slightly away from $\nu = 1$. For $\nu > 1$ 
      the ground state prior to photoexcitation already 
     contains a small number of 
   skyrmions. Provided their density is small, 
     the skyrmions can be considered as noninteracting. 
   (We do not consider filling factors further away 
     from $\nu = 1$, where interactions between  
     skyrmions can give rise to the formation 
     of a Skyrme crystal~\cite{Cote}.)
     The initial state prior to recombination
    contains an additional skyrmion and a valence hole.
    One skyrmion recombines with the hole, leaving
     behind enough spin flips to satisfy the spin selection
    rule. The final state 
     also contains the remaining skyrmions. Since only
     one skyrmion is involved in the recombination, 
   the PL spectrum is similar to that at $\nu = 1$. 

   The PL spectrum at $\nu < 1$ is qualitatively
   different from that at $\nu \geq 1$. Consider
    the filling factor at which the ground state before   
   photoexcitation contains a single antiskyrmion. We
   denote this filling factor by $\nu = 1^{-}$. The initial 
    state prior to recombination consists of a filled spin-up 
    LLL and a valence hole. 
    Since the initial state contains no spin-down electrons, 
   the RCP line is missing from the PL spectrum. 
    The final state after LCP recombination has a quasihole 
    in the spin-up LLL.\@ Because of the spin selection rule, 
      the excitation left in the final state cannot 
     be a large-spin antiskyrmion. Thus we expect no 
      PL signature of antiskyrmions at $\nu < 1$.  

   We calculate the energy of the LCP line at $\nu = 1^{-}$
     by evaluating Eq.~(\ref{eq:moments}) with $n = 0,1$
    in the initial state $|i_{-}\rangle = \prod_{m = 0}^{\infty}
     e^{\dagger}_{m \uparrow} h^{\dagger}_{0,+} |{\rm vac}\rangle$. 
    The LCP line 
   at $\nu = 1^{-}$ is shifted down in energy from the LCP line 
     at $\nu = 1$. The red shift {\em does not} depend on $g$
     when the initial state at $\nu = 1$ consists of a quasielectron
       and a valence hole. The red shift {\em does} depend on $g$
     when the initial state at $\nu = 1$ consists of a skyrmion 
       and a valence hole. The red shift can be understood by 
      comparing the final states after LCP recombination at 
     $\nu = 1^{-}$ and $\nu = 1$~\cite{Muzykantskii}. 
       While the final state at 
       $\nu = 1^{-}$ contains a free
   quasihole, the final state at $\nu = 1$ contains
    a quasihole bound to a skyrmion. The red shift is
   the binding energy of the skyrmion-quasihole pair. 
   For $\nu < 1^{-}$, the initial state contains a 
   small density of antiskyrmions in addition to the valence 
   hole. We argue that because of the Coulomb repulsion, and
   the compressibility of the electron system at $\nu < 1$, 
  the valence hole avoids the antiskyrmions and recombines 
    with a spin-up electron, leaving a quasihole in the final 
   state. Thus a small density of antiskyrmions does not
  affect the recombination. The PL spectrum is similar to 
     that at $\nu = 1^{-}$. 

    We now discuss the PL spectrum in the
  absence of disorder. For a 
    translationally invariant system, a skyrmion-hole pair
      has the center-of-mass momentum {\bf k} as a
   good quantum number~\cite{Portengen}. By analogy
    with a magnetoexciton in the LLL~\cite{Lozovik}, 
     we expect the dispersion of a skyrmion-hole pair to  
      have an absolute minimum at ${\bf k} = 0$. The initial 
    state in the absence of disorder is then a skyrmion-hole 
     pair with ${\bf k} = 0$. Because of the spin selection rule, 
     the final state contains $|S_{z} + \frac{1}{2}\sigma|$ 
    spin waves with total momentum ${\bf k} = 0$. When 
       $|S_{z} + \frac{1}{2} \sigma| > 1$ there is a continuum 
     of final states. To estimate the mean luminescence 
       energy, we calculate the energy of the transition to  
     the final state with $|S_{z} + \frac{1}{2} \sigma|$ 
       spin waves, each of momentum ${\bf k} = 0$. We expect 
      this final state to have a large oscillator strength. 
       By Larmor's theorem, the energy of this final state 
       is $-g_{e} \mu_{B} B |S_{z} + \frac{1}{2} \sigma|$. 
       Neglecting the binding energy of the skyrmion-hole 
      pair ($d = \infty$), we find                   
\begin{equation}
  \langle \omega \rangle = E_{g} + \frac{e B}{2 \mu c} 
     + E_{\rm sk} - S_{z} g_{e} \mu_{B}B .
\end{equation}
    Here $E_{\rm sk}$ is the energy of the skyrmion. 
      The red shift of the LCP line is $E_{\rm sk} -
     S_{z} g_{e} \mu_{B} B + (\pi/2)^{1/2} e^{2}/\epsilon \ell$.
   For a spin texture with radius $\lambda > 0$, the mean 
     luminescence energy and the red shift depend on $g$. 
    For $\lambda \gg \ell$ the explicit $g$ dependence 
     can be obtained using $S_{z} = -\pi(\lambda/\ell)$ 
    and Eq.~(7) of Ref.~\onlinecite{Sondhi}. For a spin
      texture with zero radius the mean luminescence energy 
      and the red shift do not depend on $g$. This can be 
     seen by setting $E_{\rm sk} = -\frac{1}{2} g_{e} \mu_{B}B$ 
    and $S_{z} = -\frac{1}{2}$.     
                   
  Polarization-resolved PL spectra from 
   GaAs-AlGaAs heterojunctions and one-side 
     doped quantum wells were reported in 
   Ref.~\onlinecite{Goldberg}.
  The ground subband emission from heterojunctions
   and wide quantum wells shows a LCP line and a RCP line 
  on the low-field side of $\nu = 1$, and a LCP line 
     on the high-field side of $\nu = 1$.
  The LCP line on the high-field side is shifted down in 
   energy from the LCP line on the low-field side by 
    $0.4$~meV in wide quantum wells and by $2$~meV
    in single heterojunctions.    
       Unpolarized PL spectra 
    from GaAs-AlGaAs heterojunctions reported in 
     Ref.~\onlinecite{Turberfield} show two lines on the 
    low-field side and one line on the high-field side 
      of $\nu = 1$. The lower-energy line on the low-field 
    side is red shifted by $1$~meV as the field increases 
     through $\nu = 1$. 

   The observations of Refs.~\onlinecite{Turberfield,Goldberg}  
   are consistent with our theoretical description of 
  luminescence near $\nu = 1$. The calculated red shifts are
    larger than the observed red shifts, partly due to our 
     neglect of the finite extent of the $z$ wavefunctions.
    The observation of a red shift as such provides no clue 
    about the nature of the negatively charged 
   excitations at $\nu = 1$. This is because a red shift is 
    expected both for skyrmions and for quasielectrons.
 It is the $g$\/-factor dependence of the red shift that
  allows the distinction to be made. Two methods of 
  varying $g$ have been employed in recent 
  magnetotransport measurements. In 
     Ref.~\onlinecite{Schmeller} $g$ was varied by tilting
 the total magnetic field $B_{\rm tot}$ away from the 
  normal to the heterojunction plane, while keeping the perpendicular
   field $B_{\perp}$ (and hence the filling factor) constant.
   While the electron spin couples to $B_{\rm tot}$, the
  orbital motion is determined by $B_{\perp}$. As a 
  result, the effective $g$ factor is $g_{e}/\cos \theta$,
   where $\theta$ is the tilting angle. An additional 
    $\theta$ dependence of the PL energies
  arises from the deformation of the $z$ wavefunctions by the parallel
  component $B_{\parallel}$ of the magnetic field.
       The $\theta$ dependence of $\langle \omega \rangle$ 
    may therefore be less suitable as a PL signature of 
      skyrmions. The red shift 
   remains unaffected by $B_{\parallel}$ because the 
  $z$ wavefunctions are deformed equally on either 
  side of $\nu = 1$. 
 Figure~\ref{fig:shift} shows the variation of the red
   shift with tilting angle for a GaAs-AlGaAs heterojunction 
     with $n_{s} = 10^{11}$~${\rm cm}^{-2}$. Tilting the field 
   by $60^{\rm o}$ increases the red shift by $0.6$~meV.\@
  Such an increase should be readily observable experimentally.  

   In Ref.~\onlinecite{Holmes} 
  $g$ was varied by applying hydrostatic 
    pressure. The applied pressure reduces the magnitude 
   of $g_{e}$ through the mixing of the GaAs conduction 
    band and the spin-orbit split valence band.  
  Reference~\onlinecite{Holmes} claims $g_{e} = -0.43 + 0.020\,p$ 
   for pressures $p$ from $0$ to $10$~kbar. 
   The applied pressure also changes other  
 bandstructure parameters. 
      However, as these parameters change
   equally on either side of $\nu = 1$, the red shift
 acquires no additional pressure dependence. 
    Figure~\ref{fig:shift}
  shows the variation of the red shift with applied
   pressure for a GaAs-AlGaAs heterojunction with 
   $n_{s} = 10^{11}$~${\rm cm}^{-2}$. Applying a 
     pressure of $10$~kbar reduces the red shift 
    by $0.4$~meV.\@   
   The problem of the decrease of $n_{s}$ with applied 
 pressure~\cite{Holmes} may be overcome by reducing 
   the field (keeping $\nu = 1$ fixed) and dividing the  
  red shift by $e^{2}/\epsilon \ell$ to correct for the
    pressure dependence of $\ell$. The division by 
 $e^{2}/\epsilon \ell$ also corrects for the small 
 pressure dependence of $\epsilon$. 
    
  We thank Andrew Turberfield, Robin Nicholas, Luis Brey, 
   Carlos Tejedor, and Lu Sham for helpful discussions. 
   This work was supported by EPSRC Grant No.\ GR/K 15619.

\begin{figure}
\caption{\label{fig:mean}
   Mean energy of the RCP and LCP 
 luminescence lines as a function of the $g$ factor,
  for various values of the separation $d$
   between the electron and hole planes.  
    Parameter values are for a GaAs-AlGaAs 
  heterojunction with $n_{s} = 10^{11}$ ${\rm cm}^{-2}$. 
    The gap value $E_{g} = 1509$ meV was taken
      from Ref.~\protect\onlinecite{Turberfield}. 
  Curves are displaced for clarity. 
    Inset: energy levels between which RCP and
    LCP transitions occur.}    
\end{figure}

\begin{figure}
\caption{\label{fig:shift}
  Variation of the red shift of the LCP line 
    with tilting angle $\theta$ and 
  applied hydrostatic pressure $p$. 
  The separation between the electron and hole
      planes is $d = 400$ \AA. Remaining
  parameter values as in Fig.~\protect\ref{fig:mean}. 
  Inset: dependence of the RCP and LCP 
    luminescence energies on the magnetic field.} 
\end{figure}


\begin{thebibliography}{99}

\bibitem{Sondhi}
S. L. Sondhi, A. Karlhede, and S. A. Kivelson,
Phys.\ Rev.\ B, {\bf 47}, 16 419 (1993).

\bibitem{Fertig}
H. A. Fertig, L. Brey, R. C\^{o}t\'{e}, and
A. H. MacDonald, Phys.\ Rev.\ B, {\bf 50}, 11 018 (1994).

\bibitem{Barrett} 
S. E. Barrett {\em et al.},
Phys.\ Rev.\ Lett.\ {\bf 74}. 5112 (1995).

\bibitem{Schmeller}
A. Schmeller, J. P. Eisenstein, L. N. Pfeiffer, and 
K. W. West, Phys.\ Rev.\ Lett.\ {\bf 75}, 4290 (1995).

\bibitem{Aifer}
E. H. Aifer, B. B. Goldberg, and D. A. Broido,
Phys.\ Rev.\ Lett.\ {\bf 76}, 680 (1996). 

\bibitem{Turberfield}
A. J. Turberfield {\em et al.},
Phys.\ Rev.\ Lett.\ {\bf 65}, 637 (1990).

\bibitem{Goldberg}
B. B. Goldberg {\em et al.}
Surf.\ Sci.\ {\bf 263}, 9 (1992).

\bibitem{Buhmann}
H. Buhmann {\em et al.},
Phys.\ Rev.\ Lett.\ {\bf 66}, 926 (1991).

\bibitem{Birman}
N. A. Viet and J. L. Birman,
Phys.\ Rev.\ B {\bf 51}, 14337 (1995).

\bibitem{Cooper}
N. R. Cooper and D. B. Chklovskii,
eprint cond-mat/9607002. 

\bibitem{Brey}
L. Brey (unpublished).

\bibitem{Bethe}
H. A. Bethe and R. Jackiw,
{\em Intermediate Quantum Mechanics},
(Benjamin/Cummings, Menlo Park, CA, 1986),
Chap.\ 11.

\bibitem{Rashba}
V. M. Apalkov and E. I. Rashba,
Pis'ma Zh.\ Eksp.\ Teor.\ Fiz.\ {\bf 53}, 420 (1991) 
[JETP Lett.\ {\bf 53}, 442 (1991)].

\bibitem{Cote}
L. Brey, H. Fertig, R. C\^{o}t\'{e}, and A. H. MacDonald,
Phys.\ Rev.\ Lett.\ {\bf 75}, 2562 (1995).

\bibitem{Muzykantskii}
B. A. Muzykantski\u{\i},
Zh.\ Eksp.\ Teor.\ Fiz.\ {\bf 101}, 1084 (1992) 
[Sov.\ Phys.\ JETP {\bf 74}, 897 (1992)].

\bibitem{Portengen}
T. Portengen (unpublished).

\bibitem{Lozovik}
I. V. Lerner and Yu.\ V. Lozovik,
Zh.\ Eksp.\ Teor.\ Fiz.\ {\bf 78}, 1167 (1978)
[Sov.\ Phys.\ JETP {\bf 51}, 588 (1980)].

\bibitem{Holmes}
S. Holmes {\em et al.},
Semicond.\ Sci.\ Technol.\  {\bf 9}, 1549 (1995).

\end{thebibliography}
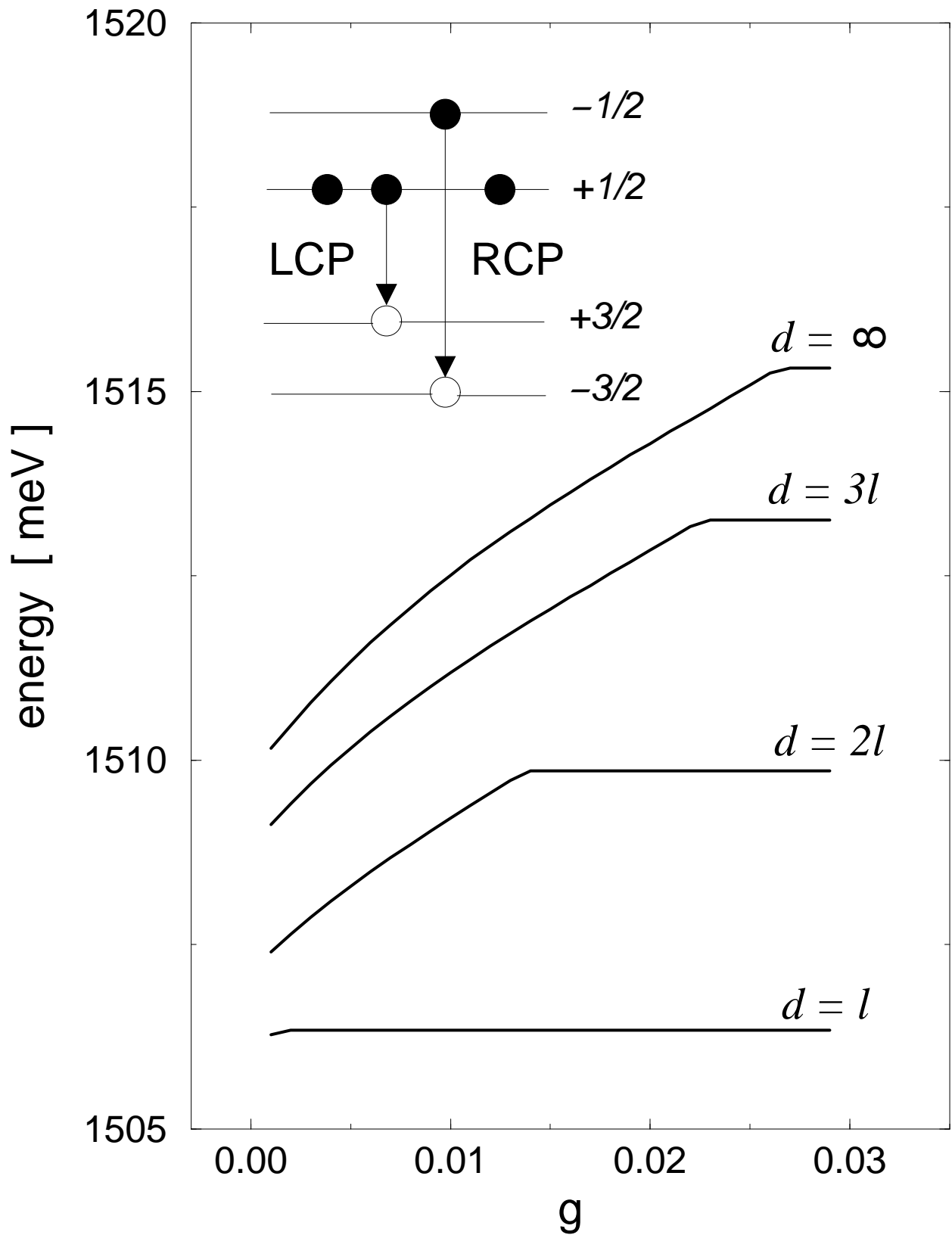
\end{document}